# Type II Perfect Absorption and Amplification Modes with Controllable Bandwidth in PT-Symmetric/Traditional Bragg Grating Combined Structures

C. Y. Huang,[1] R. Zhang,[1] J. L. Han,[2] J. Zheng,[1] and J. Q. Xu[1*]

[1] *Key Laboratory for Laser Plasmas (Ministry of Education) and Department of Physics and Astronomy, Shanghai Jiao Tong University, Shanghai 200240, China*

[2] *State Key Laboratory of Silicon Materials, Department of Material Science and Engineering, Zhejiang University, Hangzhou 310027, China*

We reveal previously unnoticed Type II perfect absorption and amplification modes of optical scattering system. These modes, in contrast to the counterparts in recent works [Phys. Rev. A 82, 031801 (2010); Phys. Rev. Lett. 106, 093902 (2011).] which could be referred as Type I modes, appear at a continuous region along the real frequency axis with any frequency. The Type II modes can be demonstrated in the PT-symmetric/traditional Bragg grating combined structures. A series of exotic nonreciprocal absorption and amplification behaviours are observed in the combined structures, making them have potential for versatile devices acting simultaneously as a perfect absorber, an amplifier, and an ultra-narrowband filter. Based on the properties of Type II modes, we also propose structures with controllable perfect absorption and amplification bandwidth at any single or multiple wavelengths.

A laser refers to a system which spontaneously emits coherent electromagnetic radiation when the pumping level exceeds its threshold [1]. As the time-reversed counterpart to laser emission, the device called coherent perfect absorber (CPA) can perfectly trap all the incoming coherent light under right conditions and convert it to some form of internal energy [2]. Instead of embedding a gain medium into the cavity as a normal laser, the CPA can be realized by adding a ``loss medium'', such as Si wafer, to the same resonator [3]. This physical phenomenon is inherently linked to a real-frequency zero of scattering matrix (S matrix). CPA may bring potential applications in integrated optics and new type of photonic devices. With utilizing a Parity-Time (PT) symmetric medium, a device can function simultaneously as a CPA and a laser at threshold, which is corresponding to a pole of the S matrix coinciding with a zero at real axis of complex k plane [4,5]. In optics, PT symmetry demands that the complex refractive index of the structures satisfy $n(z) = n^*(-z)$ [6]. This class of structures has stayed in the limelight of scientific research for last decade, mainly because of a variety of extraordinary phenomena appeared at the exceptional point. Recently, Z. Lin et al. theoretically proposed the unidirectional invisible grating induced by PT-symmetric periodic structures near the spontaneous breaking point [7]. At this breaking point, the reflection from one end is diminished while it is amplified from the other. L. Feng et al. experimentally realized a chip-size optical metamaterial device possessing unidirectional reflectionless [8].

In this Letter we uncover a set of nontrivial perfect absorption and amplification modes of optical scattering system, which are different from the mostly researched ones as mentioned in Ref. [2-5]. If the modes, which support conventional laser and CPA, are classified as Type I modes, we would like to refer what we present as the Type II perfect absorption and amplification modes in optical scattering system. A possible way to synthesize the devices that work on the Type II modes is combining PT-symmetric grating (active grating, periodic gain and loss) [7] and traditional Bragg grating (passive grating, involving no gain or loss) together. The combined structures exhibit a series of exotic nonreciprocal absorption and amplification behaviors. Within integrated photonics, a nonreciprocal device plays an important role in optical information processing. To date, various unidirectional transmission devices have been developed [9,10]. As opposed to the previous notions of nonreciprocal, we achieve unidirectional absorption and amplification via the joint effect of the PT-symmetry breaking, ``self-interference'', and Bragg periodic structure. Even more exciting is that the absorption and amplification bandwidths can be controlled due to the properties of Type II modes, which makes these combined structures more intriguing and practical for the usage like optical sensor, thermal detector, and other broadband devices.

For simplicity, let's consider a one dimensional scattering system, whose electric field envelope $E(z)$ of frequency $\omega$ obeys the Helmholtz equation:

$$\frac{d^2E(z)}{dz^2} + k^2 n(z) E(z) = 0$$

where $k = \omega/c$ and $n(z)$ is the complex refractive index which varies along the $z$ axis. If we donate $A_0$ and $A_L$ ($B_0$ and $B_L$) as complex amplitudes of the forward (backward) propagating waves at position z=0 and z=L of the system, respectively, the solution of Eq.1 can be written as $E(z) = A_0 \exp(ikz) + B_0 \exp(-ikz)$ for z<0 and $E(z) = A_L \exp(ikz) + B_L \exp(-ikz)$ for z>L. Based on the well-known transfer matrix method, they can be related by a 2×2 matrix $M$ as following:

$$\begin{bmatrix} A_L \\ B_L \end{bmatrix} = \begin{bmatrix} M_{11} & M_{12} \\ M_{21} & M_{22} \end{bmatrix} \begin{bmatrix} A_0 \\ B_0 \end{bmatrix}$$

Letting the boundary conditions $A_0 = 1$ and $B_L = 0$ ($A_0 = 0$ and $B_L = 1$), the transmission and reflection amplitudes for left (right) injection wave can be deduced as $T_f = |t|^2 = |A_L|^2 = \left|\frac{M_{11}M_{22} - M_{12}M_{21}}{M_{22}}\right|^2$ and $R_b = |r_b|^2 = |B_0|^2 = \left|\frac{-M_{21}}{M_{22}}\right|^2$ ($T_b = |t|^2 = |B_0|^2 = \left|\frac{1}{M_{22}}\right|^2$ and $R_f = |r_f|^2 = |A_L|^2 = \left|\frac{M_{12}}{M_{22}}\right|^2$), respectively. To better illustrate the fundamental differences between the Type I and II of laser and perfect absorption mode within semi-classical theory, we also write down the scattering matrix (S matrix):

$$S = \begin{bmatrix} r_b & t_f \\ t_b & r_f \end{bmatrix} = \frac{1}{M_{22}} \begin{bmatrix} -M_{21} & 1 \\ 1 & M_{12} \end{bmatrix}$$

Perfect absorption and laser emission require a device work at the zeros and poles of S matrix, respectively. In the case of conventional laser and recently proposed CPA, the above conditions are fulfilled via setting $M_{11} = 0$ and $M_{22} = 0$ [4]. When a zero and a pole of the S matrix coincide at a real frequency, perfect absorption and lasing can occur at a single device, which has been termed a "CPA-Laser". However, we also can reach $|\det(S)| \to 0$ by letting $M_{22} \to \pm\infty$ and keep other elements finite, while we can get $|\det(S)| \to \infty$ by making $\frac{M_{12}}{M_{22}}$ or $\frac{M_{21}}{M_{22}} \to \pm\infty$. These are Type II perfect absorption mode and amplification mode in optical scattering system. For the purpose of realizing the perfect absorption and amplification mode simultaneously just like the CPA-Laser working under threshold, the above conditions are simplified to $M_{22} \to \pm\infty, \frac{M_{21}}{M_{22}} \to \pm\infty$, and other two elements remain finite. The origin of the Type II perfect absorption mode and laser mode make them distinct from the previously proposed ones, endowing the devices working at the Type II modes with different propreties.

In order to demonstrate the novelty, we now move to some specific examples. Generally, it's difficult to directly achieve Type II modes from traditional optical elements. The recently introduced PT-symmetric grating, combined with traditional Bragg grating, paves the way to realizing Type II perfect absorption and amplification mode simultaneously. The proposed structure, as shown in Fig. 1, is constructed by concatenating the reflective end of the active grating with a passive grating. The refractive index distributions of active grating and passive grating are $n_{active} = n_0 + n_1 \cos(2\pi z/\Lambda) + i n_2 \sin(2\pi z/\Lambda)$ and $n_{passive} = n_0 + n_1 \cos(2\pi z/\Lambda)$, respectively. The elements of transfer matrix are given by [11,12]

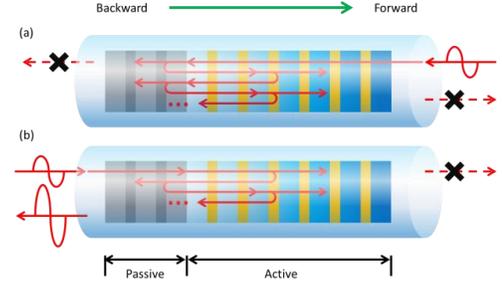

Fig.1 Schematic for the passive (grey) and active (blue with yellow strips) grating combined structures. The darker blue side donates the nonreflective end of the active grating. The solid arrows indicate existence of beams, and the dash arrows indicate the absence of beams. (a) The wave entering from right will be trapped inside the structure. (b) The wave entering from left experiences amplified reflection without transmission.

$$M_{11} = \left[\cosh(\gamma L) + i\frac{\sigma}{\gamma}\sinh(\gamma L)\right]$$

$$M_{12} = i\frac{\kappa_{12}}{\gamma}\sinh(\gamma L)$$

$$M_{21} = -i\frac{\kappa_{21}}{\gamma}\sinh(\gamma L)$$

$$M_{22} = \left[\cosh(\gamma L) - i\frac{\sigma}{\gamma}\sinh(\gamma L)\right]$$

where $\gamma = \sqrt{\kappa_{12}\kappa_{21} - \sigma^2}$ and $\sigma = \frac{n_0(\omega - \omega_0)}{c}$. For the passive grating, $\kappa_{12} = \kappa_{21} = \kappa_n = \omega_B \Delta n/(2c)$, while for the active grating, $\kappa_{12} = \kappa_n - \kappa_\alpha$, $\kappa_{21} = \kappa_n + \kappa_\alpha$. Note that the coupling coefficient $\kappa_{12}$ can be cancelled out when $\kappa_n = \kappa_\alpha$ (i.e. $n_1 = n_2$) and thus the "PT-symmetric breaking point" is reached, which is what we focus here. Then the transfer matrix of the system can be calculated as $M_{combined} = M_{active} M_{passive}$, where $M_{active}$ and $M_{passive}$ are the transfer matrix of active and passive grating. For the passive grating, $|M_{12}^{passive}|$, $|M_{21}^{passive}|$, and $|M_{22}^{passive}|$ increase fast but share almost a same trajectory when it is lengthened, whereas for the active grating these 3 elements show some inconformity but $|M_{12}^{active}| \equiv 0$ and $|M_{22}^{active}| \equiv 1$. Thus when they combined together, the asymmetry of $|M_{ij}^{active}|$ ($ij = 12, 21, 22$) break the reciprocal behavior of $|M_{ij}^{passive}|$, on the other hand, the large values of $|M_{ij}^{passive}|$ "pull up" one or two elements among $|M_{ij}^{active}|$. As a result, the poles and zeros of S matrix can be got close to. Fig. 2 shows the eigenvalues of S matrices associating with passive grating,

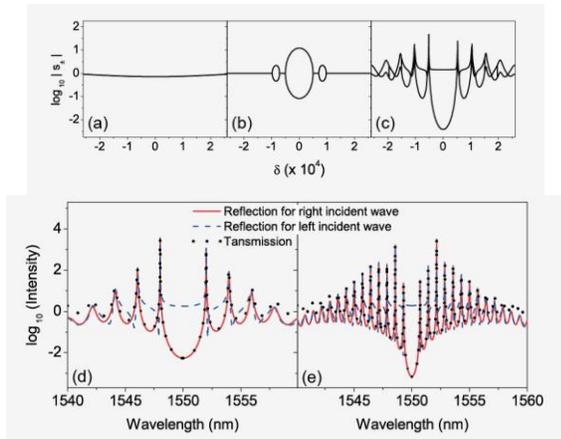

Fig. 2 Semilog plot of S matrix eigenvalues $log_{10}|s_\pm|$ for (a) passive grating, (b) active grating, and (c) combined structure as a function of δ. (d)(e)Reflection and transmission spectra for left (blue dash, black dot) and right (red solid, black dot) incident wave. We have employed the material of the Si-on-insulator (SOI) platform with $n_0 = 2.69$, $n_1 = n_2 = 0.15$ from Ref. [8], which is compatible with conventional complementary-metal-oxide-semiconductor (CMOS) technology. For other parameter values $\Lambda = 1550$ nm, (a)(c)-(e) $L_{passive} = 28.9$ μm, which corresponds to 50% transmission and 50% reflection at 1550 nm; (b)-(d) $L_{active} = 200$ μm; (e) $L_{active} = 600$ μm.

active grating, and combined structures as $\delta = (\omega - \omega_0)/c$ varies. Unlike passive grating and active grating, the combined structure always stays in the "symmetry breaking" state.

The resulting reflection and transmission spectra are displayed in Fig. 2(d)(e). When a signal with relative phase of 0 is injected from the right side of the combined structure (see Fig. 1(a)), a remarkable absorption dip is observed around the Bragg reference wavelength (over 99.9% of the incident signal is captured by the structure without transmission or reflection), which exactly implies the Type II perfect absorption mode. In distinguish to CPA [2,3], this behaviour stems from the interplay of PT-symmetry breaking, ``self-interference'', and Bragg periodic structure. As the initial signal incident into the structure, it can reach the passive grating without any change. Then a portion of the signal is reflected back to active grating, which will be introduced additional phase $\pi/2$ by the passive grating, and another portion proceeds. The reflected portion will propagate forward and simultaneously produce a magnified backward signal, which is re-injected into passive grating with another phase $\pi/2$. Through numerous times of such process, the backward (forward) signal from latter re-incident beams with relative phase of $\pi$ interfere destructively with the backward (forward) signal from the former ones with relative phase of 0, and thus the outgoing radiation have a severe reduction. One sees from Fig. 2(d)(e) that a higher order of magnitude of absorption can be achieved by increasing the length of active grating from 200 μm to 600 μm. This effect can be explained as following: owning to the destructive interference, the amplitudes of transmitted and reflected beams are dramatically diminished. In order to approach the amplitude of the previous re-incident signal, amplification is needed. Longer the active grating is, larger the re-inject signal magnified, closer the amplitude of the re-launched beam approach to the former one, and hence stronger the absorption is. Therefore, a higher order of perfect absorption can be approached through increasing active grating length. In contrast, slightly changing the length of passive grating (~20%) seems have no much effect on the performance of absorption, which proves the robust of this structure. Meanwhile, adopting higher modulation depth with appropriate material can also significantly improve the percentage of absorption, without trading the compactness of the geometry.

On the other hand, if signal is launched from left side, the transmitted signal will be absorbed, while the reflected signal amplified (see Fig. 1(b) and Fig. 2(d)(e)). Instead of corresponding to Laser mode, the photons in this scenario live in the Type II amplification mode, which is the action below the lasing threshold. In Ref. [4], the ``PT-symmetric laser absorber'', which is constructed on the Type I modes, can act as an amplifier before the threshold is reached, and as a CPA when two coherent signals are inputted. But our grating-PT-amplifier-absorber, based on the Type II modes, works in a different way: for the left side injection the signal is amplified and for the right side injection the signal is near perfectly absorbed. Hence, this intrinsic nonreciprocal absorption and amplification behaviour makes the absorber and amplifier work at the same time. The bandwidth of perfect absorption and amplification around Bragg point are ~1 nm (for intensity $< 10^{-2}$ ) and ~2 nm, respectively. Comparing to Type I modes, the second type of modes exhibit a relatively broader bandwidth, which could be a signature of Type II modes.

Moreover, Fig. 2 also indicates another interesting characteristic of the structure. The relative phases added by the gratings for the wavelengths away from the Bragg point are varies from $-\pi$ to $\pi$. For this reason, it's possible that specific wavelengths undergo cyclic constructive interference through the evolution between the active and passive gratings. As depicted in Fig. 2(a), several enhanced transmission and reflection peaks, symmetrically located around central wavelength, are as narrow as ~9 picometers (FWHM) when $L_{active} = 200$ μm, which makes the structure also have the potential to act as enhanced transmission and reflection ultra-narrow filter or amplifier for developing single-frequency laser or related devices. From the respect of S matrix, the matrix elements associating with these wavelengths lead the

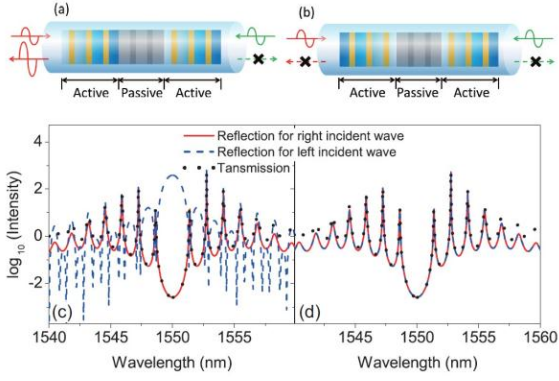

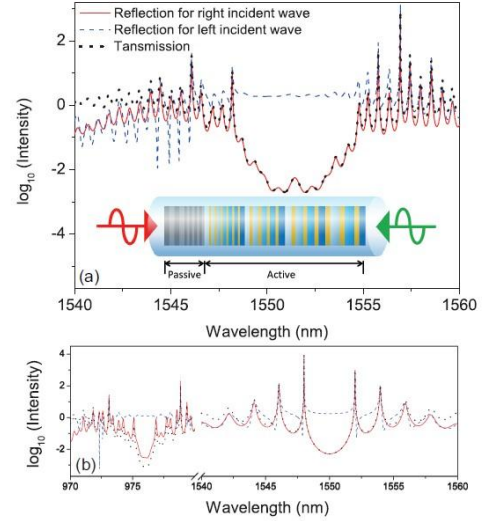

Fig. 3 (a) (b) Schematic for two extended structures. Each passive and active grating has a length of $L_{passive} = 28.9$ μm and $L_{active} = 300$ μm, respectively. The red arrows represent the left incident and reflected wave, while the green arrows represent the right incident and reflected wave. (c) (d) Reflection and transmission spectra for left (blue dash, black dot) and right (red solid, black dot) incident wave.

Fig. 4 Reflection and transmission spectra for left (blue dash, black dot) and right (red solid, black dot) incident wave. (a) Bandwidths are broadened; (b) nonreciprocal absorption and amplification occur at two bands. Inset, schematic for the combined structure with chirped periodic. Each passive grating has a length of $L_{passive} = 14.5$ μm, and each passive grating has a length of $L_{active} = 200$ μm.

value of $M_{22} \to 0$. Hence, they're possible solutions of Type I amplification modes.

Furthermore, two important extensions of above structure are also displayed in Fig. 3 (a) and Fig. 3 (b) for the sake of elucidating Type II modes better. Connected to the non-reflective end of another active grating with the passive grating side (see Fig. 3(a)), the value of $\frac{M_{21}}{M_{22}}$ gets closer to the infinite, leading the performance of amplification around Bragg wavelength greatly improved (see Fig. 3(c)) as compared to Fig. 2. Despite the fact that the solutions of Type II laser mode are mathematically observed from S matrix, they only can be infinitely approached instead of being reached. In other words, larger amplification is realizable, but the laser emission is unlikely to build from the noise in the medium. Supposing the third grating added to the combination with the other end (see Fig. 3(b)), the symmetric loci of the two active grating rebuild reciprocal by suppressing the value of $M_{12}$ and $M_{21}$ to finite at the same time. As a result, all the photons are in the Type II perfect absorption mode no matter which direction they come from (see Fig. 3(d)).

Another very important difference between Type I and Type II modes should be note. For the Type I modes, the exact zeros and poles of S matrix confine the solutions only to discrete points in the real axis. Once the working wavelength and complex refractive index are set, the length of the device is fixed. As for the Type II modes, the solutions are continuous along the real frequency axis in a region, and the restriction on the corresponding system's length is not as rigorous as the Type I modes, which makes its bandwidth can be easily controlled. To the aim of providing a broader bandwidth, we take a step further investigating the grating combined structures with chirped periodic. The proposed structure, schematically shown in the inset of Fig. 4, is composed by four active gratings (with Bragg wavelength of 1550 nm, 1551 nm, 1552 nm, and 1553 nm, respectively) and two passive gratings (with Bragg wavelength of 1550 nm and 1553 nm, respectively). Because PT-symmetry requires that the complex refractive index obeys $n(z) = n^*(-z)$, each active grating should keep a uniform period, whereas the passive part of the combined structure can employ an nonuniform chirp grating instead. From Fig. 4, one can see that the absorption bandwidth is increased to ~4 nm (for intensity $< 10^{-2}$), while amplification bandwidth is broadened to ~7 nm. Through altering the Bragg wavelength or the number of the gratings, the bandwidth and central wavelength can be simply tuned. In other words, the absorption and amplification bandwidth can be adjusted from 1 nm to theoretically infinite at any wavelength. Besides providing a boarder bandwidth, Type II modes also enable arbitrary tailored spectra. The structure we considered is similar but contains two active gratings and two passive gratings (both with Bragg wavelength of 1550 nm and 976 nm, respectively). It is clearly visible in Fig. 4 (b) that previously nonreciprocal absorption and amplification at 1550 nm are split to two bands around 1550 nm and 976 nm. This structure is potentially very useful as an amplifier. Note that the reflection for left incident wave (blue dash curve) is amplified slightly by the structure itself. However, the amplification could be greatly increased if we take a good use of absorbed energy input from right side. For example, doping the active grating with rare earths, such as erbium, the perfectly absorbed photon energy at 976 nm is stored as active ions on the upper laser level. When a signal at 1550 nm is incident from left, the stored energy can be extracted in the form of photocurrent and give rise to a large amplification.

In conclusion, we have discovered and theoretically investigated the second type perfect absorption and amplification modes of optical scattering system. The proprieties of the Type II modes are discussed within the experimentally feasible PT-symmetric and traditional Bragg grating combined structures. The nonreciprocal perfect absorption and amplification of these structures arise from the interaction of PT-symmetry breaking, ``self-interference'', and Bragg periodic structure. In contrast to Type I counterparts, the combined structures depending on Type II modes with nonuniform periodic can be used to obtain controllable bandwidth at any single or multiple wavelengths. These combined structures provide a novel route for designing a new class of nonreciprocal and multi-functional optical devices and extend the flexibility to control light.

This work was supported by the National Natural Science Foundation of China under Grant No. 61138006. The authors would like to thank Prof. Stefano Longhi for helpful discussions.

*Email: jqxu09@sjtu.edu.cn

[1] A. E. Siegman, Lasers (University Science Books, Mill Valley, CA, 1986).
[2] Y. D. Chong, L. Ge, H. Cao, and A. D. Stone, Phys. Rev. Lett. 105, 053901 (2010).
[3] W. Wan, Y. Chong, L. Ge, H. Noh, A. D. Stone, and H. Cao, Science 331, 889 (2011).
[4] S. Longhi, Phys. Rev. A 82, 031801(R) (2010).
[5] Y. D. Chong, L. Ge, and A. D. Stone, Phys. Rev. Lett. 106, 093902 (2011).
[6] K. G. Makris, R. El-Ganainy, , D. N. Christodoulides,and Z. H. Musslimani, Phys. Rev. Lett. 100, 103904 (2008).
[7] Z. Lin, H. Ramezani, T. Eichelkraut, T. Kottos, H. Cao, and D. N. Christodoulides, Phys. Rev. Lett. 106, 213901 (2011).
[8] L. Feng, Y.-L. Xu, W. S. Fegadolli, M.-H. Lu, J. E. B. Oliveira, V. R. Almeida, Y.-F. Chen, and A. Scherer, Nature Materials 12, 108 (2013).
[9] A. B. Khanikaev, S. H. Mousavi, G. Shvets, and Y. S. Kivshar, Phys. Rev. Lett. 105, 126804 (2010).
[10] Z. Wang, Y. D. Chong, J. D. Joannopoulos, and M. Soljacic, Phys. Rev. Lett. 100, 013905 (2008).
[11] S. Longhi, Phys. Rev. Lett. 105, 013903 (2010).
[12] M. Kulishov, J. M. Laniel, N. Belanger, J. Azana, and D. V. Plant, Opt. Express 13, 3068 (2005).